\documentclass[fleqn,10pt]{wlscirep}
\usepackage[utf8]{inputenc}
\usepackage[T1]{fontenc}
\usepackage{amsmath}
\usepackage{amsfonts}
\usepackage{amssymb}
\usepackage{amsthm}
\usepackage{float}
\usepackage{subfig}
\usepackage{graphicx}
\usepackage{placeins}
\usepackage{adjustbox}
\usepackage[capitalise, nameinlink]{cleveref}

\newcommand{\PAOFLOW}{ {\sf  PAOFLOW}}

\newcommand{\V}[1]{\vec{#1}}

\newcommand{\bk}{\mathbf{k}}

\newcommand{\bra}[1]{\langle #1 |}
\newcommand{\ket}[1]{| #1 \rangle}

\DeclareFixedFont{\ttb}{T1}{txtt}{bx}{n}{12} 
\DeclareFixedFont{\ttm}{T1}{txtt}{m}{n}{12}

\usepackage{listings}
\usepackage[most]{tcolorbox}
\usepackage{inconsolata}

\newtcblisting[auto counter]{sexylisting}[2][]{sharp corners, 
    fonttitle=\bfseries, colframe=gray, listing only, 
    listing options={basicstyle=\fontsize{9}{11}\ttfamily,language=Python}, 
    title=Listing \thetcbcounter: #2, #1}

\title{Relaxation time approximations in  PAOFLOW 2.0}

\author[1,*]{Anooja Jayaraj}
\author[2]{Ilaria Siloi}
\author[3]{Marco Fornari}
\author[1,4]{Marco Buongiorno Nardelli}
\affil[1]{Department of Physics, University of North Texas, Denton, TX 76203, USA}
\affil[2]{Department of Physics and Astronomy, University of Southern California, Los Angeles, CA  90007, USA}
\affil[3]{Department of Physics and Science of Advanced Materials Program, Central Michigan University, Mt.Pleasant, MI 48859}
\affil[4]{Santa Fe Institute, Santa Fe, NM 87501, USA}

\affil[*]{AnoojaJayaraj@my.unt.edu}

\keywords{Electronic conductivity, Relaxation Time Approximation, Thermoelectric materials}

\begin{abstract}
Regardless of its success, the constant relaxation time approximation has limited validity. Temperature and energy dependent effects are important to match experimental trends even in simple situations.
We present the implementation of relaxation time approximation models in the calculation of Boltzmann transport in \PAOFLOW\ 2.0 and apply those to model band-structures. In addition, 
using a self-consistent fitting of the model parameters to experimental conductivity data, we provide a flexible tool to extract scattering rates with high accuracy. We illustrate the approximations using simple models and then apply the method to GaAs, Si, Mg$_3$Sb$_2$, and CoSb$_3$.
\end{abstract}

\begin{document}
\maketitle
\section{Introduction}

Over the years, first-principles calculations have become a complementary tool for the experimental research aiming to discover high-performance thermoelectrics. This has greatly improved the understanding of the origin of the transport properties and has advanced optimization strategies based on electronic band-structure. \cite{witkoske2017thermoelectric,jiang2020achieving,Li2015} An important parameter often used in the screening of thermoelectric materials is the figure of merit, ZT. It is a quantitative descriptor of the materials' efficiency in converting a thermal gradient in electrical power. It is calculated using the equation,
\begin{equation}
    ZT = \frac{\sigma S^2}{\kappa}T
\end{equation}
where $\sigma$ is electrical conductivity, S is Seebeck coefficient, $\kappa$ is thermal conductivity, and T is the temperature (on an absolute scale) at which the device is operating. However, tuning ZT in order to maximize the value offers many challenges; one of which is the competing nature of Seebeck coefficient and electrical conductivity. A higher ZT would require a larger $\sigma S^2$, but the properties that favor larger $\sigma$ result in a smaller S.  A good understanding of the various transport coefficients is vital in improving the predictive ability of computational characterization of materials.
\newline
The most standard approach in calculating the transport coefficient is to use the semi-classical Boltzmann theory within the constant relaxation time approximation (CRTA). In most cases, this has been used without much analysis. Though it works for certain systems, recent research has shown that the CRTA have resulted in wrong predictions, missing vital information to understand the transport properties.\cite{xu2014first,zhou2018large,sun2015large} Moreover, the introduction of an arbitrary constant of relaxation time severely limits the predictive capabilities of first-principles calculations.  In this work, we investigate how the CRTA affects the electronic transport coefficients properties of well known TE materials by considering different scattering models that include energy and temperature dependence. Though methods exist for {\it ab initio} calculation of electron-phonon relaxation times, \cite{zhou2021perturbo,ponce2016epw} the calculation of the electron-phonon matrix requires extremely dense $\textbf{k}$ and $\textbf{q}$ point meshes. Therefore, prohibitively high computational costs make these techniques of limited practicality, especially when aiming to data driven high-throughput approaches. We chose to combine accurately interpolated band-structures and simplified mathematical models of the scattering phenomena in order to explore the consequences on the transport coefficients beyond CRTA and parabolic bands.
Extrinsic scattering mechanisms (impurities, grain
boundaries, alloy disorder) contribute significantly to the transport properties in a system. Indeed, these extrinsic scattering mechanisms may be often tuned during the synthesis of the system. \cite{shuai2017tuning} This raises a need to characterize the contribution of various scattering mechanism in a system in order to gain understanding of how they can be controlled to obtain optimized performaces. In this work, we introduce a self-consistent fitting of transport properties  to experimental data which will give us insight into the temperature dependence of various scattering mechanisms for specific experimental samples.
These relaxation time approximation (RTA) models are coded in our recently released  PAOFLOW package\cite{ PAOFLOW18,cerasoli2021advanced}. In this paper, we will discuss in detail the theory and implementation of the RTA models in the newest release of the software, and illustrate the automated workflow while, at the same time, documenting the influence of scattering phenomena in the band-structure of cubium, graphene, and selected materials: Si, GaAs, Mg$_3$Sb$_2$, and CoSb$_3$. 

\section{Methods}

\subsection{ PAOFLOW}\label{ PAOFLOW}

\ PAOFLOW\ is a software tool to efficiently post-process standard first-principles electronic structure plane-wave pseudopotential calculations in order to promptly compute, from interpolated band-structures and density of states, several quantities that provide insight on transport, optical, magnetic and topological properties such as  anomalous and spin Hall conductivity, magnetic circular dichroism, spin circular dichroism, and topological invariants. 
The methodology is based on the projection on pseudo-atomic orbitals (PAO)\ discussed in detail in Refs.~\citeonline{Agapito_2013_projectionsPRB,Agapito:2016jl,Agapito:2016en}.

Accurate PAO Hamiltonian matrices can be built from the direct projection of the Kohn-Sham (KS) Bloch states $\ket{\psi_{n\bk}}$ onto a chosen  basis set of fixed localized functions.
The Hamiltonian for a specific material, $ \hat{H}\left({\bf R}\right)$, is computed in real space
using atomic orbitals or pseudo atomic orbitals from the pseudopotential of any given element. The key, in this procedure, is in the mapping of the {\it ab initio} electronic structure (solved on a well converged and large plane waves basis set) into tight-binding (TB) formalism that precisely reproduces a selected number of bands of interest. The crucial quantities that measure the accuracy of the basis set are the projectabilities $p_{n\bk}=\bra{\psi_{n\bk}} \hat{P} \ket{\psi_{n\bk}} \ge 0$ ($\hat{P}$ is the operator that projects onto the space of the PAO basis set, as defined in Ref.~\citeonline{Agapito:2016jl}) which indicate  the representability of a Bloch state $\ket{\psi_{n\bk}}$ on the chosen PAO set.
Maximum projectability, $p_{n\bk}= 1$, indicates that the particular Bloch state can be perfectly represented in the chosen PAO set; contrarily, $p_{n\bk} \approx 0$ indicates that the PAO set is insufficient and should be augmented. 
Once the Bloch states with good projectabilities have been identified, the PAO Hamiltonian is constructed  as 
\begin{equation}
\hat{H}(\bk) = AEA^\dagger + \chi \left( I-A \left( A^{\dagger}A \right)^{-1}A^\dagger \right).
\label{eq:Hk16}
\end{equation}
Here $E$ is the diagonal matrix of KS eigenenergies and $A$ is the matrix of coefficients obtained from projecting the Bloch wavefunctions onto the PAO set. Since the filtering procedure introduces a null space, the parameter $\chi$ is used to shift all the unphysical solutions outside a given energy range of interest. The procedure in Eq. \ref{eq:Hk16} is recommended for most cases.

Band-structure interpolation on arbitrary Monkhorst and Pack (MP) {\bf k}-meshes for the integration in the Brillouin zone (BZ) are at the very core of the ability of \ PAOFLOW\ to provide high-precision electronic structure data. 
Indeed, the TB Hamiltonian can be Fourier transformed from real space representation to the {\bf k}-space and interpolated 
using an efficient  procedure based on a zero-padding algorithm and fast Fourier transform routines.

The same accuracy defined by the projectabilities is conserved in this process. 
The expectation values of the momentum operator, which is the main quantity in the definition of the transport coefficients, is given by

\begin{eqnarray}
{\bf p}_{nm} ({\bf k})&=&  \left < \psi_n ({\bf k}) \right |\hat{p}\left | \psi_m ({\bf k}) \right > = \\ \nonumber
&=&  \left < u_n ({\bf k}) \right |\frac{m_0}{\hbar} \V{\nabla}_{\bf k} \hat{H}({\bf k})\left | u_m ({\bf k}) \right >
\label{momentum}
\end{eqnarray}
with
\begin{equation} \label{Hamiltonian.gradient}
\V{\nabla}_{{\bf k}} \hat{H}({\bf k}) =
\sum_{\bf R} i{\bf R} \exp\left(i{\bf k}\cdot {\bf R}\right) \hat{H}\left({\bf R}\right).
\end{equation}
$\hat{H}\left({\bf R}\right)$ being the real space PAO matrix and  $\left | \psi_n ({\bf k}) \right > = \exp({-i {\bf k} \cdot {\bf r}}) \left | u_n ({\bf k}) \right >$ the Bloch's functions.\cite{pino}

\subsection{Boltzmann transport} \label{boltzmann}

In \ PAOFLOW\ the electrical conductivity is evaluated by solving the semi-classical Boltzmann equation (BTE) that describes 
the evolution of the distribution function $f$ of an electron gas under external electric field and in presence of scattering mechanisms.\cite{Parravicini:2000ud,Singh2001125,Madsen200667} 
In the so-called scattering-time approximation,  the conductivity tensor $\sigma_{ij}$ can be expressed as an integral over the first BZ:
\begin{equation} \label{boltzmann.conductivity.equation.generalized}
\sigma_{ij}= \frac{e^2}{4\pi^3} \int_{BZ} \sum_n \tau_n({\bf k})  v_n^i({\bf k})v_n^j({\bf k}) \left( -\frac{\partial f_0}{\partial E}\right)d {\bf k},
\end{equation}
where $\tau_n({\bf k}) $ is the relaxation time, $v_n^i({\bf k})$ is the {\em i}-th component of the  electron velocity corresponding to the  {\em n}-th band for each {\bf k}-point in the BZ (${\bf v}_n$ is derived by the diagonal of the momentum matrix element, Eq. \ref{momentum}),  $f_0$ is the equilibrium distribution function, and $E$ is the electron energy.

Generalizing  Eq. (\ref{boltzmann.conductivity.equation.generalized}) it is also possible to define analogue  expressions for the Seebeck coefficient $S$ and the electron contribution to thermal conductivity $\kappa_{el}$. 
Following the notation of Ref.~\citeonline{Mecholsky:2014jv}, we  introduce the { generating tensors} ${\mathcal L}_{\alpha}$ ($\alpha = 0, 1, 2$):
\begin{equation} \label{generating.tensors}
 \mathcal L_{\alpha} = \frac{1}{4\pi^3} \int \sum_n   \tau_n({\bf k}) {\bf v}_n({\bf k}){\bf v}_n({\bf k}) \left( -\frac{\partial f_0}{\partial E}\right)\left[\epsilon_n({\bf k})-\mu\right]^{\alpha}d {\bf k},
\end{equation}
where $ {\bf v}_n({\bf k}){\bf v}_n({\bf k})$ indicates the dyadic product, $\epsilon_n({\bf k})$ the band-structure, and $\mu$ is the chemical potential. The coefficients $\sigma$, $S$ and $\kappa_{el}$ can be expressed as follows:
\begin{eqnarray}
\sigma &=& e^2\mathcal L_0, \\ \nonumber
S     &=& -\frac{1}{T e} \left[\mathcal L_0\right]^{-1} \cdot \mathcal L_1, \\ \nonumber
\kappa_{el} &=& \frac{1}{T} \left(\mathcal L_2 - \mathcal L_1 \cdot \left[\mathcal L_0\right]^{-1} \cdot \mathcal L_1\right),
\end{eqnarray}
where $T$ is the temperature. 
Our formalism based on PAO-TB performs the computations of the band-velocities and avoids issues with possible band-crossing. In addition, from Eq. (\ref{boltzmann.conductivity.equation.generalized}-\ref{generating.tensors}), it is evident that the evaluation of the transport properties requires an accurate integration
over a fine grid of k-point in the BZ which becomes a trivial task using the TB representation from the PAO projections and Eq. (\ref{Hamiltonian.gradient}).

\subsection{Relaxation Time Approximation}
\label{Electronic scattering}

The most common implementations of the Boltzmann transport equations assume the scattering time $\tau$ to be a constant (CRTA). A constant $\tau$ factors out of Eq. \ref{generating.tensors} and, thus, the method returns the quantities $\sigma_0 = \sigma/\tau$ and $\kappa_{el,0} = \kappa_{el}/\tau$ (in the Seebeck coefficient $\tau$ cancels out). Clearly this is a severe approximation for a quantity that is expected to be significantly
temperature-dependent and energy-dependent (typically via a power law). A direct
estimate of the dependence of $\tau$ on energy and temperature is an important complement to any transport study, and would provide, even if at a phenomenological level, important insight into the relevant scattering mechanism present in any given system. Moreover, a direct comparison with existing experimental data would provide an extra layer of characterization for real world applications.

The approach implemented in \ PAOFLOW\ is based on the work of Jacoboni {\it et al.}~\cite{jacoboni2010theory} and recently included in the BoltzTrap\cite{boltztrap} framework by the group of V. Fiorentini.\cite{farris2018theory} They employed analytical energy-dependent expressions for the relaxation time, which were developed on the basis of known semiclassical theories and include the most important mechanisms of electron scattering by
acoustic phonons, polar-optical phonons, and charged impurities.

Acoustic phonon scattering is treated within the elastic deformation potential approach in the long-wavelength
acoustic-phonon limit, 
\begin{equation} 
\label{Acoustic scattering}
\tau_{ac}(E,T)= \frac{2\pi\hbar^{4}\rho v^{2} }{(2m^{*})^{\frac{3}{2}}k_BT D_{ac}^{2} \sqrt E}, \\
\end{equation}
where $E$ is the electron energy and T is the temperature. All other parameters are defined in Table \ref{Scattering parameters}. 

Similarly to the assumptions that were used in acoustic phonon scattering, we model optical phonon scattering with an elastic deformation potential  ($D_{op}$):

\begin{equation} 
\label{Optical scattering}
\tau_{op}(E,T)= \frac{\sqrt{2k_{B}T}\pi x_{o}\hbar^{2}\rho}{m^{{*}^{\frac{3}{2}}}D_{op}^{2}[\ N_{op}\sqrt{x+x_{o}}+(N_{op}+1)\Theta(x-x_{o})\sqrt{x-x_{o}}]\ },
\end{equation}

\begin{equation}
    N_{op} = \frac{1}{\exp\frac{{\hbar\omega_{op}}}{k_B{T}}-1},  \\
    x = \frac{E}{k_{B}T},  \\
    x_{o} = \frac{\hbar\omega_{op}}{k_{B}T}.  \\
\end{equation}
The first term in the denominator of Eq.~\ref{Optical scattering} represents the absorption of optical phonons by electrons and the second term represents the emission of the optical phonons by electrons. The probability of emission of a phonon when $E < \hbar\omega_{op}$ is zero since the electron does not have enough energy to emit the phonon and this is represented by the Heaviside step function $\Theta$ included in the second term. $N_{op}$ represents the number of optical phonons.

Polar optical scattering is modeled following Ridley:\cite{ridley1998polar}
\begin{equation}
\label{Polar optical scattering}
\tau_{pop}(E,T) = \sum_{i}\frac{Z(E,T,\omega_{i}^{l})E^{\frac{3}{2}}}{C(E,T,\omega_{i}^{l})-A(E,T,\omega_{i}^{l})-B(E,T,\omega_{i}^{l})}
\end{equation}
where the sum is over all longitudinal-optical phonons, with energy $\omega_{i}^{l}$
i ; the functions A, B, C, and Z are omitted for brevity and can be found in Appendix I.

For impurity scattering we use the Brooks-Herring approach:\cite{PhysRev.115.1107}
\begin{equation} 
\label{Impurity scattering}
\tau_{imp}(E,T) = \frac{E^{\frac{3}{2}} \sqrt{2m^{*}}4\pi\varepsilon^2} {(log(1+\frac{1}{x})-\frac{1}{1+x})\pi n_I Z_I^{2}e^{4}}
\quad \textrm{with} \quad 
x = \frac{E}{k_{B}T}.
\end{equation}

Finally, in compound semiconductors the strain induced by acoustic phonons creates a piezoelectric field. This piezoelectric scattering is  
modelled as in Ref.~\citeonline{jacoboni2010theory}.
\begin{equation} 
\label{Polar acoustic scattering}
\tau_{pac}(E,T) = \frac{\sqrt{2E}2\pi\varepsilon^2\hbar^2\rho v^2}{p^2 e^2 \sqrt{m^*} k_BT}
\times \left[1-\frac{\epsilon_{o}}{2E}\log(1+4\frac{E}{\epsilon_{o}})+\frac{1}{1+4\frac{E}{\epsilon_{o}}}\right]
\end{equation}
where $\varepsilon=\epsilon_{o}+\epsilon_{\infty}$ and the piezoelectric effect is captured by the piezoelectric constant, p. 

The global relaxation time is then obtained using Matthiessen's rule:
\begin{equation}
\label{matths}
\frac{1}{\tau_{total}(E,T)}=\frac{1}{\tau_{imp}(E,T)}+\frac{1}{\tau_{ac}(E,T)}
+\frac{1}{\tau_{op}(E,T)}+\frac{1}{\tau_{pop}(E,T)}+\frac{1}{\tau_{pac}(E,T)}.
\end{equation}

\begin{table}[H]

\centering
  \begin{tabular}{|p{0.5\textwidth}cc|}
    \hline
     Parameter & Symbol & Units \\ 
     \hline
     Mass density & $\rho$ & $kg/m^{3}$ \\ 
     Lattice constant & a & m  \\
     Low freq. dielectric constant & $\epsilon_{0}$ & -  \\
     High freq. dielectric constant & $\epsilon_{\infty}$ & -  \\
     Acoustic velocity & v & m/s \\
     Effective mass ratio & $m^{*}$ & - \\
     Acoustic deformation potential & $D_{ac}$ & eV\\
     Optical deformation potential & $D_{op}$ & eV \\
     Optical phonon energy & $\hbar\omega_{op}$ & eV\\
     Number of impurities & $n_{I}$ & $cm^{-3}$\\
     Charge on impurity & $Z_{I}$ & - \\
     Piezoelectric constant & p & $C/m^{2}$ \\
    \hline
  \end{tabular}
  \caption{Symbols and units for the scattering parameters required in various scattering models.}
  \label{Scattering parameters}
\end{table}

\section{Simple models and the parabolic band approximation}\label{smpba}

In order to quantify the improvement of a richer RTA, and to gain a better understanding how varying various parameters affect the overall transport properties of a system, we start with two simple TB models: cubium and graphene. Cubium was chosen as representative of a 3D solid with quasi-parabolic bands (near the BZ center, $\Gamma)$ and graphene for its 2D character and its linear dispersion at the Fermi level. The TB Hamiltonian for a system with two atoms per unit cell with contributions from a  single orbital is given by 

\begin{equation*}
\mathcal{H}(\textbf{k}) = 
\begin{bmatrix}
E_{g}/2 & -t\Delta_{\textbf{k}} \\
-t\Delta_{\textbf{k}}^{*} & -E_{g}/2 \\
\end{bmatrix},
\end{equation*}

where t is the first nearest-neighbor hopping parameter and $E_{g}$ defines the band gap of the band-structure. 
$
\Delta_{\textbf{k}} = \sum_{\boldsymbol{\delta}}e^{i\textbf{k}.\boldsymbol{\delta}}
$
gives the sum is over nearest neighbors. For the  cubium, because of its simple structure, the vectors for the six nearest neighbours are $\delta = a(\pm1,0,0),a(0,\pm1,0),a(0,0,\pm1)$, where $a$ is the lattice constant, so that:

\begin{align*}
\Delta_{\textbf{k}} &= e^{ik_{x}a} + e^{-ik_{x}a} + e^{ik_{y}a} +e^{-ik_{y}a} + e^{ik_{z}a} + e^{-ik_{z}a}\\
&= 2(\cos{k_{x}a}+\cos{k_{y}a}+\cos{k_{z}a}).
\end{align*}

Similarly for graphene, the vectors for the three nearest neighbors are $\delta = \frac{a}{2}(1,\sqrt{3})$, $\frac{a}{2}(1,-\sqrt{3}), -a(1,0)$ and 
\begin{align*}
\Delta_{\textbf{k}} &= e^{i\textbf{k}.\boldsymbol{\delta}_{1}} +  e^{i\textbf{k}.\boldsymbol{\delta}_{2}} + e^{i\textbf{k}.\boldsymbol{\delta}_{3}}  \\
&= e^{-ik_{x}a}\left[1 + 2e^{3ik_{x}a/2}\cos{\frac{\sqrt3k_{y}a}{2}}\right].
\end{align*}

Setting $E_{g}$ to 0 eV reproduces a graphene like band-structure where the bands show a linear dispersion at the Dirac point $K$ in the BZ.

\begin{figure}
\includegraphics[width=\textwidth]{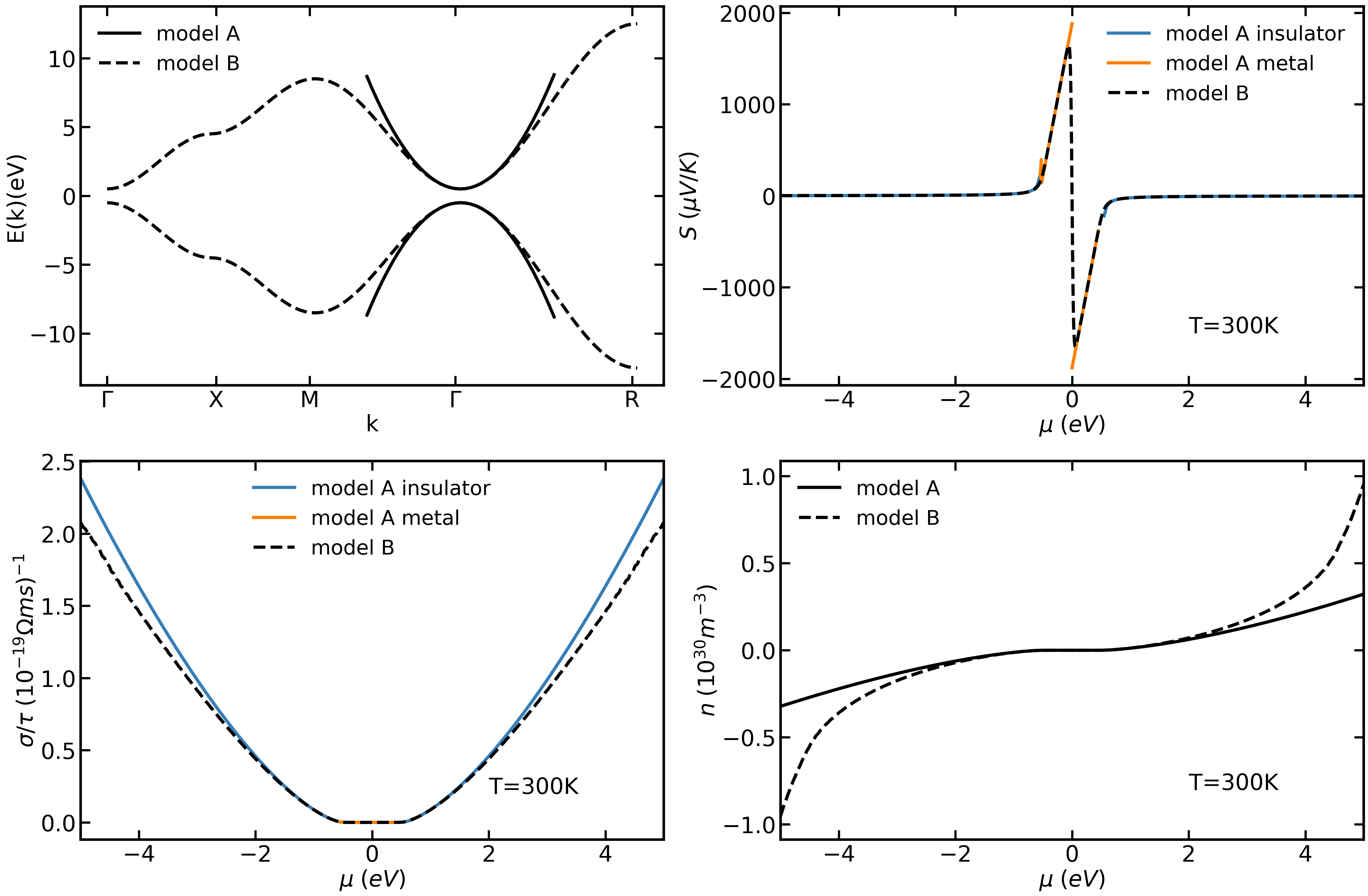}
\caption{Band-structure of a cubium with two bands (top left panel , dashed line) and a parabolic fit near $\Gamma$ point (solid line). In the top right, bottom left, and bottom right panels, the Seebeck coefficient, the conductivity,  and the carrier concentration, respectively, are reported.}
\label{fig:cubium2}
\end{figure}

Since the energy and temperature dependence of the functional for of the scattering times were obtained using a parabolic band approximation, it is useful to examine the validity of such an approximation within electronic transport. The formulas for transport coefficients for a parabolic band within the CRTA, derived in detail in Ref.~\citeonline{mecholsky2014theory}, have been compiled in Eq. \ref{analy_sig ins} - \ref{analy_carrier met}.\\
In the insulating regime for parabolic bands, when $\beta(E_{n}-\mu)\gg 1$
\begin{equation}
    \textbf[\ \boldsymbol{\sigma}\textbf]\ _{i,j} = \frac{e^{2}\tau 2^{3/2}\sqrt{m_{x}m_{y}m_{z}}}{3\pi^2\hbar^{3}m_{i}} [\ m_{n}(\mu-E_{n})]\ ^{3/2} \delta_{i,j}
    \label{analy_sig ins}
\end{equation}
\begin{equation}
    \textbf[\ \textbf S\textbf]\ _{i,j} = -m_{n}\frac{k_{B}}{2e}\left[\ 5+m_{n}2\beta(E_{n}-\mu)\right]\ \delta_{i,j}
    \label{analy_S ins}
\end{equation}
\begin{equation}
    n = \frac{\sqrt{m_{x}m_{y}m_{z}}{\exp(-\beta(E_{n}-\mu))}}{\sqrt{2}\hbar^{3}\pi^{3/2}\beta^{3/2}}
    \label{analy_carrier ins}
\end{equation}
and in the metallic regime for parabolic bands, when $\beta(E_{n}-\mu)\ll -1$
\begin{equation}
    \textbf[\ \boldsymbol{\sigma}\textbf]\ _{i,j} = \frac{e^{2}\tau 2^{3/2}\sqrt{m_{x}m_{y}m_{z}}}{3\pi^2\hbar^{3}m_{i}}[\ m_{n}(\mu-E_{n})]\ ^{3/2} \delta_{i,j}
    \label{analy_sig met}
\end{equation}
\begin{equation}
    \textbf[\ \textbf S\textbf]\ _{i,j} = -\frac{k_{B}\pi^{2}}{2e\beta(E_{n}-\mu)} \delta_{i,j}
    \label{analy_S met}
\end{equation}
\begin{equation}
    n = {\left[\ \frac{{-2(m_{x}m_{y}m_{z})^{1/3}}(E_{n}-\mu)}{3^{2/3}\hbar^{2}\pi^{4/3}}\right]\ }^{3/2}
    \label{analy_carrier met}
\end{equation}
where, $E_{n}$ represents either a band-edge minimum or maximum, and $m_{n}$ is +1 for a conduction-like band, and -1 for a valence-like band, $m_{x}, m_{y}$ and $m_{z}$ are the x, y and z components of the effective mass, $\mu$ is the chemical potential of interest, $\beta=\frac{1}{k_{B}T}$ and $n$ is the charge carrier concentration. We use the cubium model (Eq.~\ref{cubium band eqn}) and compare the transport properties to those of the parabolic bands in Figure. \ref{fig:cubium2}:

\begin{equation}
 \textnormal{model A (Parabola)} : E(\textbf{k}) = -\hbar^2|\textbf{k}|^2/2m,
\end{equation}

\begin{equation}
 \textnormal{model B (Cubium)} : E(\textbf{k}) = -6+2(cos(k_{x}a)+cos(k_{y}a)+cos(k_{z}a)),
\label{cubium band eqn}
\end{equation}

\begin{equation}
 \textnormal{model C (Graphene)} : E(\textbf{k}) = \pm2.7\sqrt{1+4\cos\left(\frac{3}{2}k_{x}a\right)\cos\left(\frac{\sqrt{3}}{2}k_{y}a\right)+4\cos^{2}\left(\frac{\sqrt{3}}{2}k_{y}a\right)} \pm 0.25.
\label{graphene band eqn}
\end{equation}

\FloatBarrier
\FloatBarrier
As apparent from the top left panel of Figure \ref{fig:cubium2}, the cubium bands start deviating from the parabolic bands at $\sim$1.5 eV. Subsequently, all the transport properties for the parabolic model and the cubium band-structure are expected to match from 0 eV to $\sim$1.5 eV as the cubium band is a good approximation of a parabolic band in this range. This is reflected in the rest of the panels of Figure \ref{fig:cubium2}. Since the parabolic band approximation is no longer valid in the cubium model after this limit, the transport properties of cubium start to deviate from those calculated using Eq. \ref{analy_sig ins} - \ref{analy_carrier met}. This deviation from parabolicity in transport properties is slightly enhanced at $\sim$4 eV due to contributions from the flat feature of the bands of cubium around the X point of the band-structure.

\begin{figure}
\includegraphics[width=\textwidth]{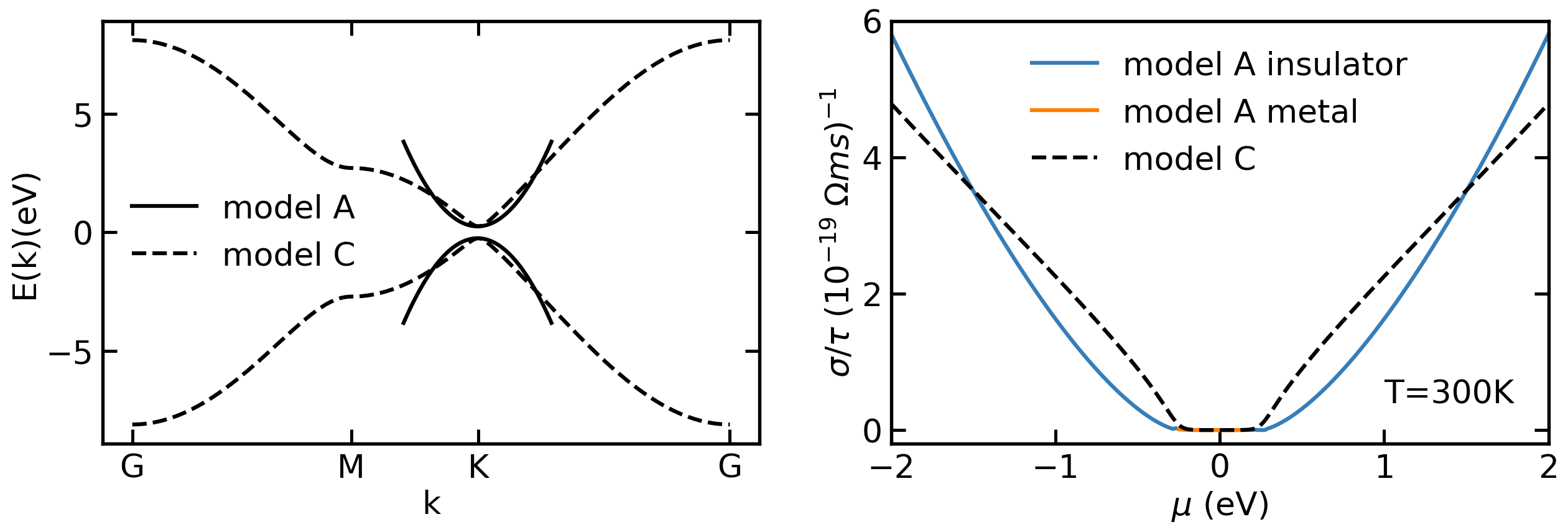}
\caption{Band-structure of a graphene model and parabolic fit at $K$ (left panel, dashed and solid line, respectively). The right panel shows the corresponding conductivities computed with Eqs.~\ref{analy_sig ins}, \ref{analy_sig met}, and  PAOFLOW.}
\label{fig:graphene}
\end{figure}

In Figure \ref{fig:graphene}, we consider a graphene like band-structure with a band gap of 0.5 eV. As expected, graphene bands do not follow a parabolic approximation, except very close to the Fermi surface. Therefore, the conductivity calculated using a parabolic band approximation is able to reproduce the conductivity calculated using BTE only extremely close to the valence and conduction band edges.  

\subsection{Simple models beyond CRTA}

Simple models allow to investigate the effect of improving the CRTA with minimal computational overload and monitor the consequences in physically transparent scenarios. Our simplified approach is conducive to an exploration of the parameters' space of the RTA models which can be then used with more realistic band-structures.
We can control the parameters of the models to enhance one or the other scattering mechanisms by varying the values of the deformation potentials, or of the relevant optical frequencies or of the sound velocity, to name a few. This can give insight into the potential design of materials with optimal properties for any given application.

\begin{figure}
\includegraphics[trim={0 0 0 2cm},clip,width=\textwidth]{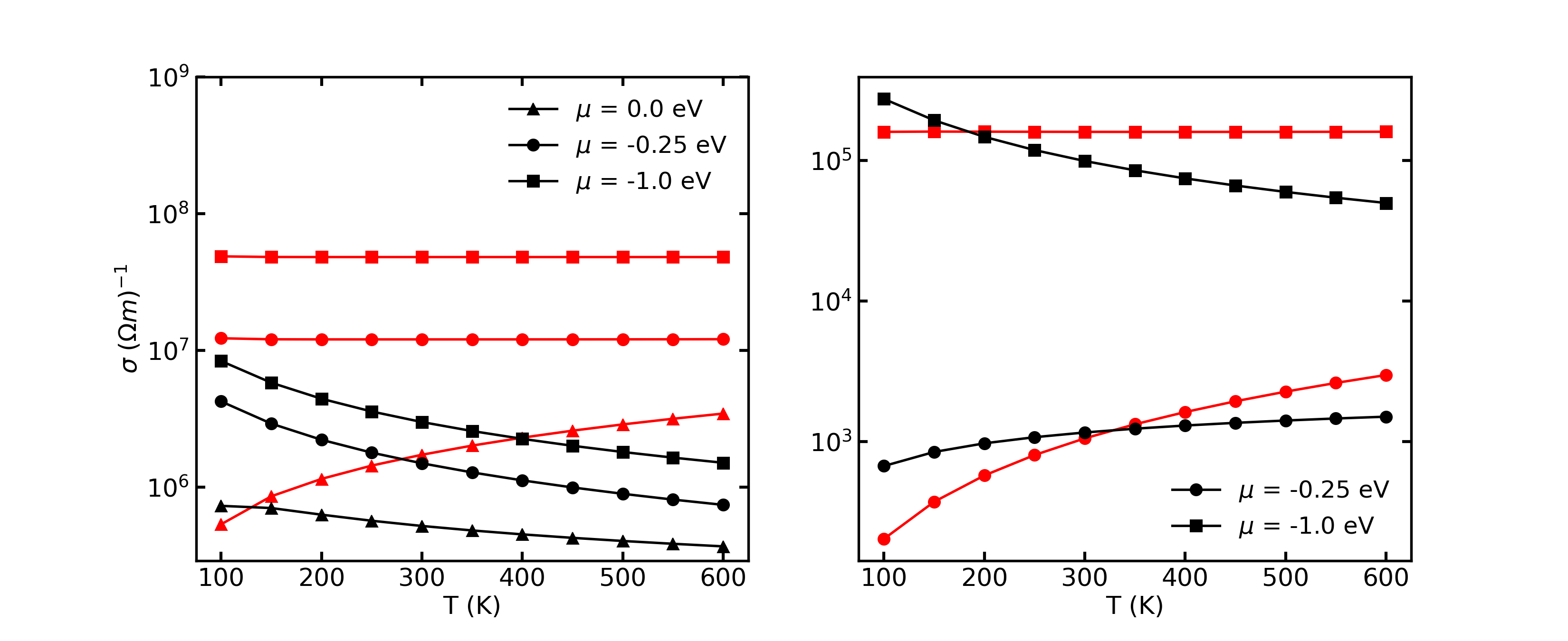}
\caption{Electronic conductivity as a function of temperature at various chemical potentials (representing the doping level, referred to the middle of the gap). The left panel corresponds to graphene (with zero band gap) and the right panel corresponds to cubium with a band gap of 0.5 eV.
The red lines denote the conductivity calculated using the CRTA whereas the black lines correspond to the conductivity calculated using the RTA; different markers correspond to different chemical potentials as in the legends.
}
\label{graphene cubium }
\end{figure}

A first observation involves the comparison between the conductivity calculated within CRTA and other other RTA models. We present results for graphene and cubium, Figure~\ref{graphene cubium }. 
Experimentally, the distinction between semiconducting and metallic behavior (including the case of heavy doping) is understood in term of the temperature dependence of the conductivity. Samples whose conductivity increases with temperature are semiconducting and samples whose conductivity decreases with temperature are metallic. In semiconductors, the increase in the number of charge carriers prevails over the reduction of the relaxation time; in metals, the reduction of $\tau$ induces the reduction of $\sigma$.

In the case of graphene (a quasi-linear dispersion), the CRTA ($\tau = 10^{-14}$ s) provides temperature-independent conductivity in the metallic case and monotonically increasing conductivity in the semiconducting regime (the position of the chemical potential wrt the band edge determine the regime from the electronic structure point of view): only the variation of the carrier density due to temperature is captured in the calculation (Figure~\ref{graphene cubium }, left panel, red lines). When applied to a two-band cubium with a forbidden energy gap of 0.5 eV, the same phenomenology is obtained in the CRTA (Figure~\ref{graphene cubium }, right panel, red lines).
Let's consider a specific RTA model constructed using 
$[{D_{ac}:1,\rho:1e3,v:1e3,ms:1,D_{op}:5e10,h\omega_{lo}:0.01}]$ and optical phonon and acoustic phonon scattering mechanisms (see Section~\ref{Electronic scattering}). In graphene, the chosen RTA model introduce dissipation phenomena that shorten $\tau$ as the temperature increases: this induce a reduction in conductivity. In the cubium model, we recover the experimental evidence of decreasing conductivity as function of temperature in the heavily-doped (metallic case) and increasing conductivity as function of temperture in the semiconducting regime.

\section{Relaxation time models and experimental conductivity}

The scattering models described in Section~\ref{Electronic scattering} have limited validity when extrinsic effects such as size of dopants or impurities, edge effects etc. can significantly affect the scattering time. The electronic conductivity calculated using the relaxation times from the models, however, provide a framework for comparison to experiments and provide insight on  the wide variations in experimental conditions and doping. We propose a modified Mathiessen's rule as: 

\begin{equation}
\label{matths_modify}
\frac{1}{\tau_{total}(E,T)}=\frac{a_{imp}(T)}{\tau_{imp}(E,T)}+\frac{a_{ac}(T)}{\tau_{ac}(E,T)}
+\frac{a_{op}(T)}{\tau_{op}(E,T)}+\frac{a_{pop}(T)}{\tau_{pop}(E,T)}+\frac{a_{pac}(T)}{\tau_{pac}(E,T)}.
\end{equation}

In the above equation, $a_{imp}$, $a_{ac}$, $a_{op}$, $a_{pop}$ and $a_{pac}$ are correcting functions that are fitted to reproduce the experimental conductivity.  The fitting procedure uses the sequential least squares programming (SLSQP) \cite{nocedal2006sequential} method, which allows for constrained non linear optimization of the fitting functions. 
A more detailed discussion of the fitting procedure is presented in Appendix 2.
To demonstrate the effectiveness of this approach and its implementation in the \ PAOFLOW\ package, we present results for four prototypical systems: GaAs, Si, Mg$_3$Sb$_2$ and CoSb$_3$. 

\subsection{Computational details and implementation in \ PAOFLOW.} \label{comp_details}

The calculation of the Boltzmann transport with the modified RTA models is implemented in \ PAOFLOW\ and follow a standard algorithmic flow. \ PAOFLOW\ requires a few basic calculations performed with the Quantum ESPRESSO (QE) package. \cite{Giannozzi:2009p1507,giannozzi2017advanced} The first (self-consistent) run generates a converged electronic density and Kohn-Sham (KS) potential on an appropriate Monkhorst and Pack (MP) {\bf k}-point mesh. The second (non self-consistent) one evaluates eigenvalues and eigenfunctions on a larger MP mesh and often for an increased number of bands. After these preliminary steps \ PAOFLOW's most fundamental procedure is the construction of accurate PAO Hamiltonians following the theory outlined in Section \ref{ PAOFLOW}.

The density functional theory (DFT) calculations for Si were performed using a Local Density Approximation (LDA). A kinetic energy cut off of 18 Ry (180 Ry cut off for the charge density) and a 12 $\times$ 12 $\times$ 12 {Monkhorst–Pack \bf k}-point mesh were used for the non self-consistent calculation. This was further increased to a 150 $\times$ 150 $\times$ 150 grid using  PAOFLOW's Fourier interpolation method in order to accurately integrate transport tensors.

The DFT calculations for GaAs were performed using an generalized gradient approximation (GGA) functional in the parametrization of Perdew, Burke, and Ernzerhof (PBE) . Projector Augmented Wavefunctions (PAW) were used to treat the ion-electron interactions. A kinetic energy cut off of 60 Ry (600 Ry cut off for the charge density) and a 16 $\times$ 16 $\times$ 16 {Monkhorst–Pack \bf k}-point mesh were used. This was further increased to a 100 $\times$ 100 $\times$ 100 grid using  PAOFLOW.

The DFT calculations for $\text{Mg}_{3}\text{Sb}_{2}$ were performed using a GGA functional in the parametrization of PBE. PAW were used to treat the ion-electron interactions. A kinetic energy cut off of 45 Ry (450 Ry cut off for the charge density) and a 24 $\times$ 24 $\times$ 18 {Monkhorst–Pack \bf k}-point mesh were used. This was further increased to a 96 $\times$ 96 $\times$ 72 grid using  PAOFLOW.

The DFT calculations for $\text{CoSb}_{3}$ were performed using a local density approximation (LDA) functional. A kinetic energy cut off of 45 Ry (450 Ry cut off for the charge density) and a 10 $\times$ 10 $\times$ 10 {Monkhorst–Pack \bf k}-point mesh were used. This was further increased to a 100 $\times$ 100 $\times$ 100 grid using  PAOFLOW.
The pseudopotentials for all the atomic species were obtained from pslibrary1.0.0.\cite{DALCORSO2014337}

The experimental parameters used in the calculations of the relaxation times for different systems have been listed in Table \ref{Scattering parameter value}. The energy, E is taken from the original DFT Hamiltonian processed by \ PAOFLOW. The calculations are done for user defined ranges of temperature T. An example workflow is discussed in Appendix 3.

\begin{table}[H]
\centering
  \begin{tabular}{|cccccc|}
    \hline
     Symbol & Units & $\text{Mg}_{3}\text{Sb}_{2}$ & GaAs & Si & $\text{CoSb}_{3}$\\ 
     \hline
     $\rho$ & kg/m$^{3}$ & 3.9$\times$10$^{3}$ & 5.3$\times$10$^{3}$ & 2.3$\times$10$^{3}$ & 7.8$\times$10$^{3}$\\ 
     a & m  & 8.7$\times$10$^{-10}$ & 5.6$\times$10$^{-10}$ & 5.4$\times$10$^{-10}$ & 9.1$\times$10$^{-10}$\\
     $\epsilon_{0}$ & - & 26.7 & 13.5 & 11.7 & 33.5\\
     $\epsilon_{\infty}$ & - & 14.2 & 11.6 & - & 25.6\\
     v & m/s & 2.7$\times$10$^{3}$ & 5.2$\times$10$^{3}$ & 6.6$\times$10$^{3}$ & 3.3$\times$10$^{3}$\\
     m$^{*}$ & - & 0.3 & 0.7 & 0.29 & 3\\
     D$_{ac}$ & eV & 6.5 & 7 & 9.5 & 5\\
     D$_{op}$ & eV & - & - & 8$\times$10$^{10}$ & 1$\times$10$^{11}$\\
     $\hbar\omega_{LO}$ & eV & [0.0205,0.0248,0.031] & [0.03536] & - & [0.0264]\\
     p & C/m$^{2}$ & - & 0.16 & - & -\\
     
    \hline 
  \end{tabular}
\caption{Symbols and units of the parameters to be input in calculation of scattering      	models. The values for Si and GaAs are obtained from Ref.~\citeonline{jacoboni2010theory}, for $\text{Mg}_3\text{Sb}_{2}$ from Ref.~\citeonline{farris2018theory} and for $\text{CoSb}_{3}$ from Refs.~\citeonline{tang2015convergence,arushanov1997transport}.}
  \label{Scattering parameter value} 

\end{table}

\subsection{GaAs}
\label{GaAs}

The scattering models were implemented on n-type GaAs for two different carrier concentrations, $3.5\times10^{17}\;cm^{-3}$ and $7.7\times10^{18}\;cm^{-3}$ and was used to calculate conductivities. 

\begin{figure}
\includegraphics[width=\textwidth]{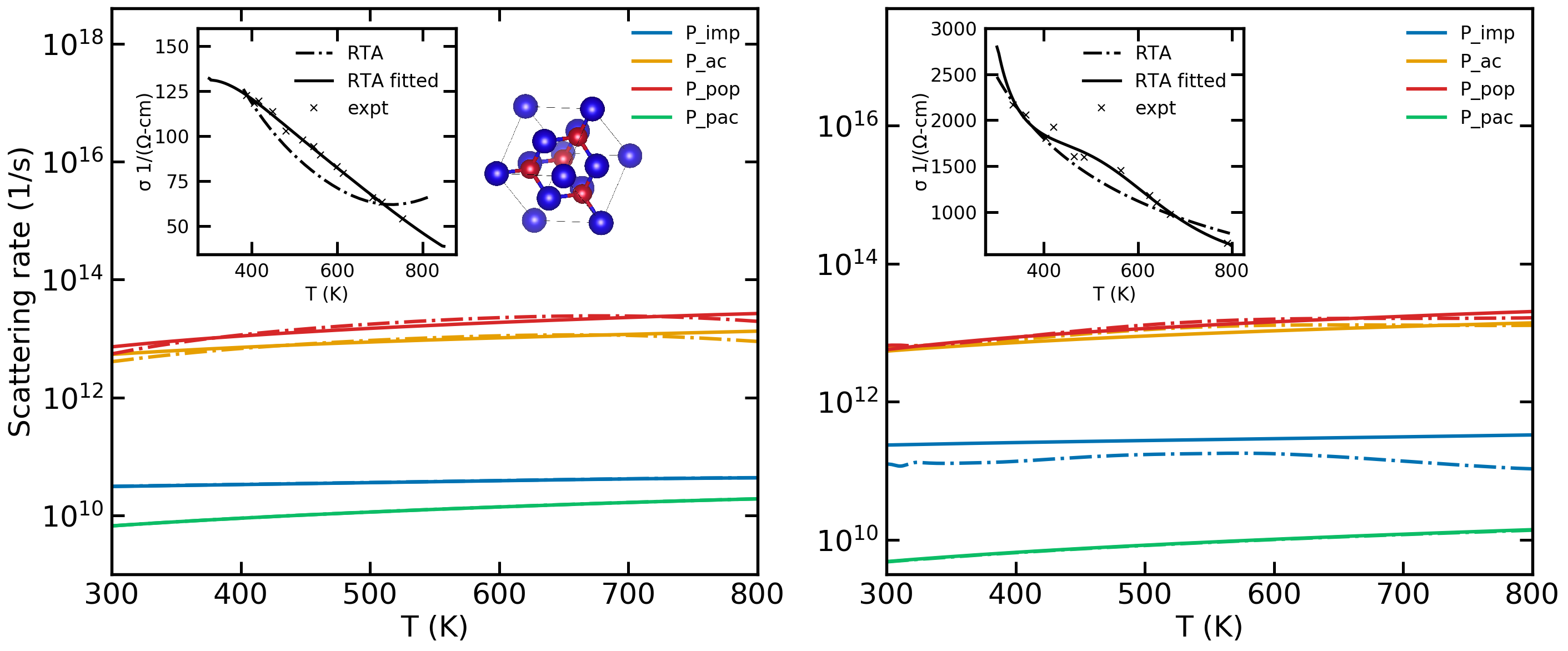}
\caption{A comparison of the scattering rates in GaAs obtained using the original scattering models (labelled RTA) and those obtained using the fitting procedure (labelled RTA fitted) for samples of two different doping concentrations. The left panel corresponds to an n-type sample with a doping concentration of $3.5\times10^{17}cm^{-3}$ while the right panel corresponds to an n-type doping of $7.7\times10^{18}cm^{-3}$. The inset shows the electrical conductivity calculated using both the original and the fitted scattering rates.The experimental data have been obtained from Ref.~\citeonline{amith1965electron}.}
\label{GaAs optimization 1}
\end{figure}

The scattering rates as a function of temperature are depicted by the solid lines in Figure \ref{GaAs optimization 1}. The dominant scattering mechanisms were determined from Refs.~\citeonline{amith1965electron} and \citeonline{lee1979electrical}. The calculated conductivities were then compared to the experimental values as shown in the inset of Figure \ref{GaAs optimization 1}. The fitting procedure is performed as well and the scattering rates obtained as a result of the fitting procedure are shown by the dashed lines in Figure \ref{GaAs optimization 1} (a) and (b). The fitting procedure produces negligible changes to the scattering rates, signifying that the original models themselves represent the scattering rates in GaAs well. 
\\Ref.~\citeonline{amith1965electron}, from which the experimental data have been obtained, uses analysis of their Seebeck and Hall coefficient data and notes that the relative weight of polar scattering increases with increasing temperature, whereas the contribution of impurity scattering decreases with increasing temperature. This is confirmed by our results as well.

\subsection{Si}
\begin{figure}
\includegraphics[width=\textwidth]{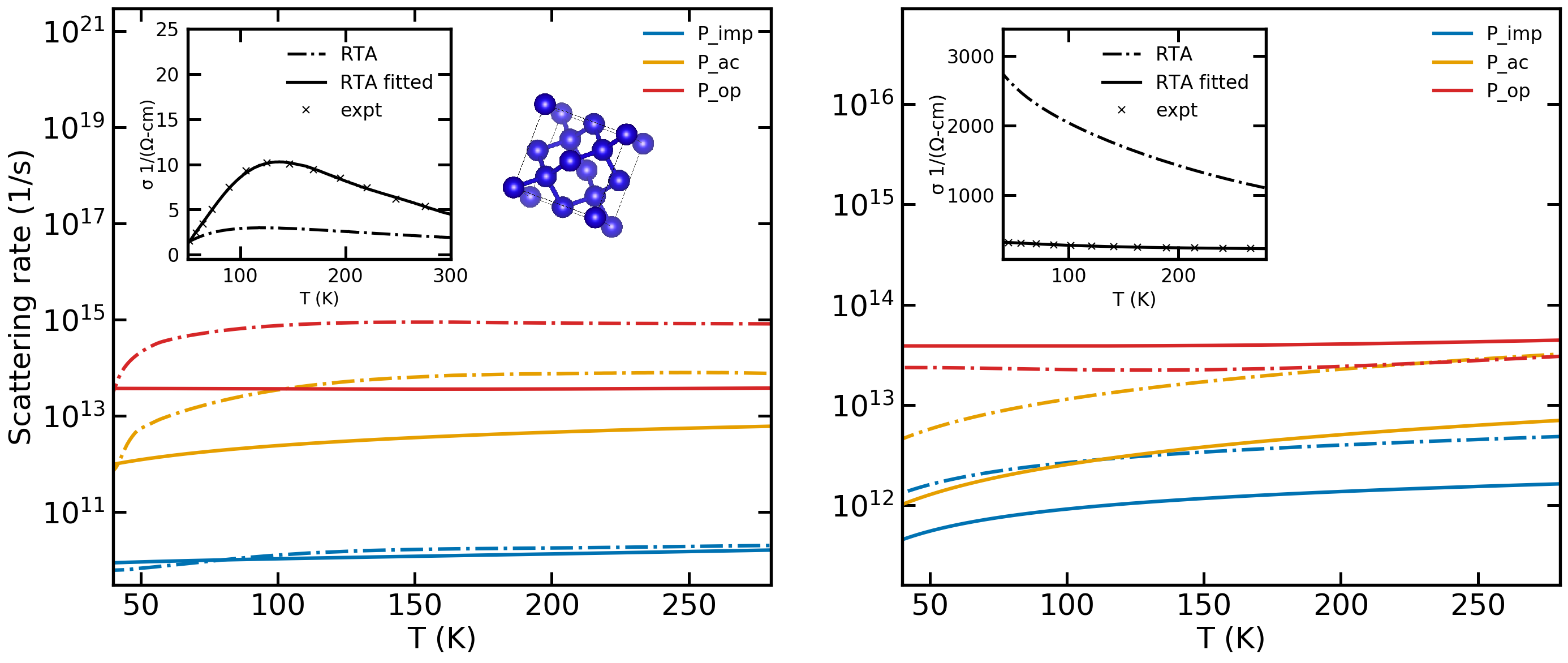}
\caption{Conductivities and scattering rates for the different Si samples in a low temperature regime. The left panel shows the data for a sample with n-type doping with a carrier concentration of 2.8x10$^{16}$cm$^{-3}$ while the right panel shows the data for a sample with n-type doping with a carrier concentration of 1.7x10$^{19}$cm$^{-3}$. The inset shows the goodness of fit of theoretical electrical conductivity to experiments resulting from the fitting procedure as well as the electrical conductivity calculated using the original scattering models}
\label{Si optimization 1}
\end{figure}

Electronic conductivities were calculated for intrinsic Si, and n-type Si with doping concentrations of $2.8\times10^{16}$ cm$^{-3}$ and $1.7\times10^{19}$ cm$^{-3}$. Similar to GaAs, the solid lines in Figure \ref{Si optimization 1} represent the scattering rates calculated from the original models while the dashed lines represent the scattering rates obtained using the modified Mathiessen's rule. It is clear from the results that the original scattering rates require significant modification for the calculated conductivities to match experiments. 
\\Low temperature experimental data for transport properties in Si were obtained from Ref.~\citeonline{weber1991transport}. There, the authors discuss the electrical conductivity of the heavily doped sample (n =1.7x10$^{19}$cm$^{-3}$) being weakly temperature dependent due to the large number of impurity atoms forming an impurity band. They state that in contrast, the weakly doped n=2.8x10$^{16}$cm$^{-3}$ exhibits an exponential behavior of electrical conductivity due to the freeze out of impurities at low temperatures. This behavior is fully captured by our models as well. In panel (a) of Figure \ref{Si optimization 1}, for n=2.8x10$^{16}$cm$^{-3}$ the fitting procedure produces negligible change to the scattering rate due to impurity scattering in order to match experimental data. However, in panel 2 of Figure \ref{Si optimization 1}. For n=1.7x10$^{19}$cm$^{-3}$ we see the contribution of impurities to the overall scattering rate is underestimated by the original models, but is rectified by our fitting procedure which significantly increases the scattering rate due to impurities. The dominant scattering mechanisms are the optical and acoustic phonons scattering whose values require significant correction by the fitting procedure.

\subsection{Mg3Sb2}

\begin{figure}
\includegraphics[width=\textwidth]{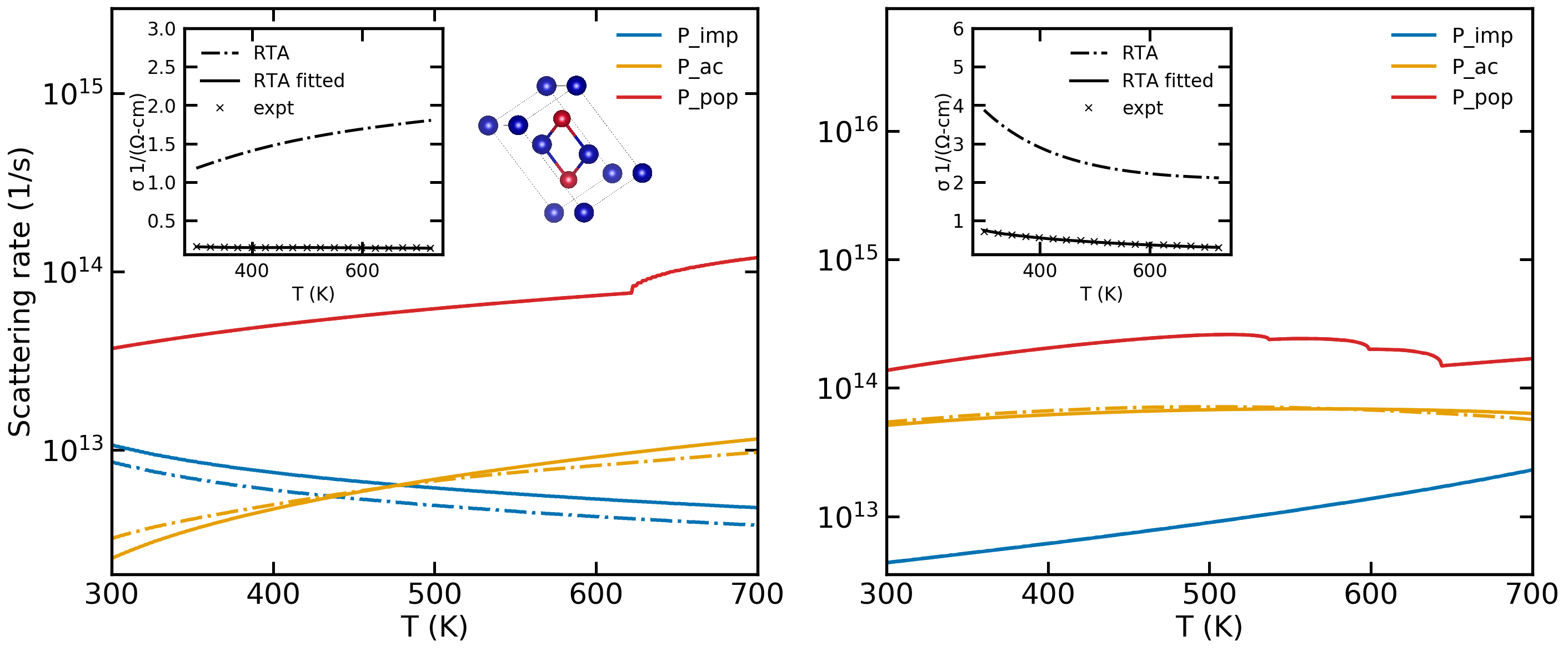}

\caption{A comparison of the scattering rates in Mg$_3$Sb$_2$ obtained using the original scattering models (RTA) and those obtained using the fitting procedure (RTA$_\text{fitted}$). The left panel corresponds to an n-type sample with a doping concentration of $3.6\times10^{18}cm^{-3}$ while the right panel corresponds to an n-type doping of $3.6\times10^{19}cm^{-3}$. The inset shows the electrical conductivity calculated using the respective scattering rates.The experimental data have been obtained from Ref.~\citeonline{farris2018theory}}
\label{Mg3Sb2}
\end{figure}

$\text{Mg}_{3}\text{Sb}_{2}$, a well studied thermoelectric was chosen as a test system in order to verify our implementation of the various scattering models as well as the fitting procedure. The effect of scattering rates on transport properties in $\text{Mg}_{3}\text{Sb}_{2}$ and comparison to experiments have been extensively carried out in Ref.~\citeonline{farris2018theory}. Their results show that transport properties calculated with scattering models are in good agreement with experiments. This is confirmed by our implementation of scattering models and the subsequent fitting procedure. As shown in Figure \ref{Mg3Sb2}, the fitting procedure produces negligible modifications to the original scattering models in order to match calculated electrical conductivity to experimental data.

\subsection{CoSb3}
\label{CoSb3}
The conductivity for the p-type samples of $\text{CoSb}_{3}$ seem better represented by the scattering models than that of those for n-type samples. As seen from the fitting parameters for various samples, the ones for p-type samples are lower than those for n-type samples. The scattering rates for n-type sample require significant corrections at higher temperatures in order to reproduce experimental results.
\\CoSb$_3$, a well known thermoelectric, is studied over a wide range of temperatures. Caillat $et$ $al$.\cite{caillat1996properties} analyzed their experimental mobility data for p-type samples and suggests that the dominant scattering mechanism, at least below 500 K is acoustic phonon scattering since the mobility followed a T$^{-3/2}$ behaviour. However, our models seem to suggest that the dominant scattering mechanism is optical phonon scattering for p-type samples and depending on the doping levels, maybe acoustic phonon scattering or optical phonon scattering for n-type samples. This inconsistency is also noted by Y. Kajikawa\cite{kajikawa2014analysis} who carried out analysis of p-type CoSb$_3$ within a two-valence and two-conduction band model.

\begin{figure}
\includegraphics[width=\textwidth]{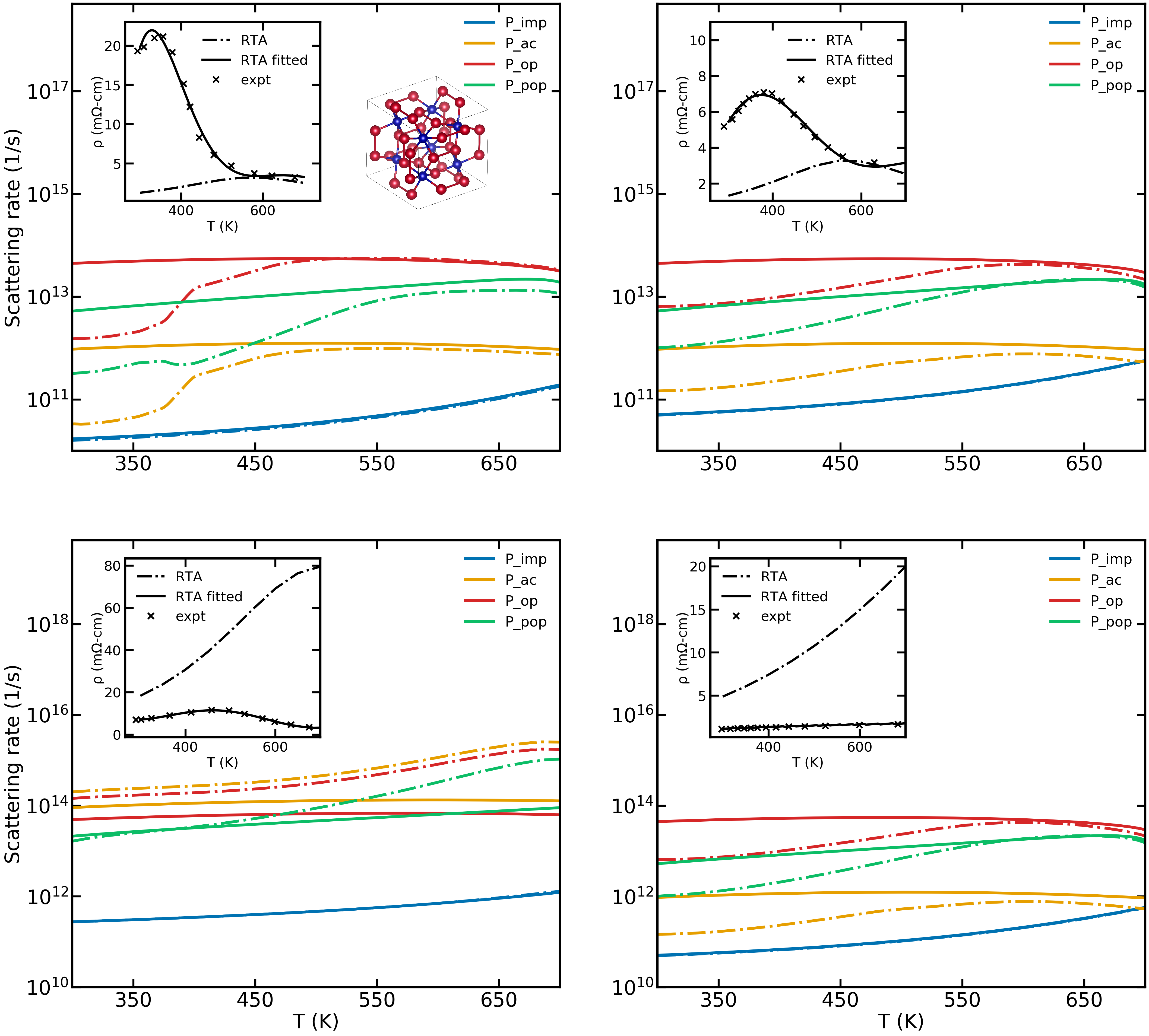}

\caption{A comparison of the scattering rates in CoSb$_3$ obtained using the original scattering models (RTA) and those obtained using the fitting procedure (RTA$_\text{fitted}$) are shown. The top left panel corresponds to a p-type sample with a doping concentration of $1.2\times10^{17}cm^{-3}$ while the top right panel corresponds to a p-type doping of $4.4\times10^{17}cm^{-3}$. The bottom left panel corresponds to an n-type sample with a doping concentration of $152\times10^{17}cm^{-3}$ while the bottom right panel corresponds to an n-type doping of $1380\times10^{17}cm^{-3}$. The inset shows the electrical resistivity calculated using the both the original and fitted scattering rates. The experimental data have been obtained from Ref.~\citeonline{caillat1996properties}}.
\label{CoSb3 optimization 1}
\end{figure}

\section{Conclusion}

We have implemented relaxation time models that allow calculation of conductivities beyond the constant relaxation time approximations and are able to provide reasonable agreement to experimental conductivities in various systems. Moreover, unlike the CRTA, it allows for a quantitative and qualitative description of the scattering mechanisms themselves. We introduce an automated self consistent fitting procedure that allows one to see how various the base scattering models need to be tuned in order to reproduce experimental conductivities. This is highly advantageous in determining sample specific scattering properties which is beyond the scope of the base models. 

\section{Acknowledgments} 
The authors wish to acknowledge the support and computational guidance from Frank Cerasoli and Andrew Supka; fruitful scientific discussions with Jagoda S\l awi\'{n}ska, Priya Gopal, and Nicholas Mecholsky; and the computational resources provided by the High Performance Computing Center at the University of North Texas and the Texas Advanced Computing Center at the University of Texas, Austin. 
The authors acknowledge partial support by DOE (DE-SC0019432).

\section{Appendix I}
Formulas of the functions entering the definition of the optical phonon scattering time:
\begin{equation} 
\begin{aligned}
\label{Polar optical scattering A}
A(E,T,\omega_{op}) =n(\omega_{op}+1)\frac{f_{0}(E+\hbar\omega_{op})}{f_{0}(E)}[(2E+\hbar\omega_{op})\\
&sinh^{-1}(\frac{E}{\hbar\omega_{op}})^{\frac{1}{2}}-[E(E+\hbar\omega_{op})]^\frac{1}{2}]
\end{aligned}
\end{equation}
\begin{equation} 
\begin{aligned}
\label{Polar optical scattering B}
B(E,T,\omega_{op}) =\theta(E-\hbar\omega_{op})n(\omega_{op})\frac{f_{0}(E-\hbar\omega_{op})}{f_{0}(E)}\\
&[(2E-\hbar\omega_{op})cosh^{-1}(\frac{E}{\hbar\omega_{op}})^{\frac{1}{2}}\\
&-[E(E-\hbar\omega_{op})]^\frac{1}{2}]
\end{aligned}
\end{equation}
\begin{equation} 
\begin{aligned}
\label{Polar optical scattering C}
C(E,T,\omega_{op}) =2E[n(\omega_{op}+1)\frac{f_{0}(E+\hbar\omega_{op})}{f_{0}(E)}\\
&sinh^{-1}(\frac{E}{\hbar\omega_{op}})^{\frac{1}{2}}+\theta(E-\hbar\omega_{op})n(\omega_{op})\\
&\frac{f_{0}(E-\hbar\omega_{op})}{f_{0}(E)}cosh^{-1}(\frac{E}{\hbar\omega_{op}})^{\frac{1}{2}}]
\end{aligned}
\end{equation}
\begin{equation} 
\begin{aligned}
\label{Polar optical scattering Z}
Z(\omega_{op}) =\frac{2}{W_{0}(\hbar \omega_{op})^\frac{1}{2}},
\,W_{0}(\omega_{op})=\frac{e^{2}\sqrt{2m^{*}\omega_{op}}\varepsilon^{-1}}{4\pi\hbar^{\frac{3}{2}}}
\end{aligned}
\end{equation}

\section{Appendix 2}
\label{Appendix 2}
We implemented a fitting procedure as follows. We extract experimental data for the electronic conductivity (or resistivity) as a function of temperature for the system of interest. Since the experimental data points may be few and far apart, this data is interpolated using a polynomial fit of a degree that best fits the available data. This will allow for a smooth fitting procedure without nonphysical discontinuities in the final scattering times.
The fitting proceeds to minimize the distance $f(T)$ between the experimental conductivity curve and the calculated conductivity curve by varying the fitting functions ($a_{im}$, $a_{ac}$, $a_{op}$...), where it is assumed that the functions vary with temperature: 

\begin{equation}
\label{minimize}
f(T) = \sum_{i=1}^{N}\frac{(\sigma_{exp}^{i}(T)-\sigma_{pao}^{i}(T))^{2}}{N},
\end{equation}

where $\sigma_{exp}$ and $\sigma_{pao}$ are the experimental and the theoretical values of conductivity respectively, with N being total number of data points taken into consideration.
Since the conductivity curves  often span a wide range of temperatures and show different behaviors in different ranges, we introduce a moving overlapping bin fitting procedure to capture these changes and obtain smooth fits. This is done by splitting both the experimental and calculated data into overlapping bins with at least $f$ data points, $f$ being the number of fitting parameters. Therefore the bins for the experimental conductivities with four fitting parameters would be \\ $[\ \sigma_{exp}(T_{1}), \sigma_{exp}(T_{2}),\sigma_{exp}(T_{3}),\sigma_{exp}(T_{4}) ]\  , [\ \sigma_{exp}(T_{2}), \sigma_{exp}(T_{3}),\sigma_{exp}(T_{4}),\sigma_{exp}(T_{5})]\ $ and so on, covering the entire temperature range. The calculated conductivities are split into similar bins and the distance between the curves represented by the calculated conductivity bin and the experimental conductivity bin is minimized using the following steps.\\
 Set the initial guess for the fitting parameters ($a_{imp}$, $a_{ac}$, $a_{op}$...) of the first bin to ones so that the starting point for the of the fitting procedure coincides with the familiar Matthiessen's rule (Eq.(\ref{matths})). The total scattering time as a function of $E(\textbf{k},T)$ obtained from Eq.(\ref{matths_modify}) is plugged into Eq.(\ref{generating.tensors}) in order to calculate the electrical conductivity. The fitting parameters are allowed to vary until the distance between the first bin of the experimental curve and the first bin of the calculated curve is minimized. We will refer to these set of fitting parameters as the converged fitting parameters and the relaxation times calculated using these converged fitting parameters as converged relaxation times.\\
 This minimization procedure is carried out for every bin. In order to speed up the fitting procedure, the converged fitting parameters of the previous bin is taken to be the initial guess for the fitting parameters of every subsequent bin. This also removes any nonphysical jumps in relaxation times between the bins.\\
 The converged relaxation times obtained for every temperature in every bin is combined. Since the bins overlap in temperature, every temperature will have multiple relaxation times as well. These are averaged to obtain one converged relaxation time per temperature to obtain the final relaxation time vs temperature plot.

\section{Appendix 3}

Listing of the main.py from example 10 of the \ PAOFLOW\ package where we construct the workflow to reproduce the results of GaAs with a doping concentration of $3.5\times10^{17}\;cm^{-3}$ described in Section \ref{GaAs}.
For a complete discussion of method and attributes of the  PAOFLOW class see \cite{cerasoli2021advanced}. 
Once the interpolated Hamiltonian is built and the gradient and momenta are computed, various transport properties can be calculated for any system. Since the systems are typically doped, {\bf doping\_conc} argument  is passed to the {\textit {doping}} routine which computes the chemical potential required to fix the doping concentration for various temperatures. The energy and temperature dependent scattering models are defined with the {\tt TauModel} class. The built in models include acoustic, optical, polar acoustic, polar optical and ionized impurity scattering. These built-in models only require the specification of empirical constants.  The empirical constants required for any selected built-in models are passed into a python dictionary  as the {\bf tau\_dict} argument of the {\tt transport} routine.\PAOFLOW\ also allows the user to define scattering models that can be passed directly to the {\bf scattering\_channels} argument as demonstrated by the {\bf acoustic\_model} function.  The variable {\bf channels} then uses 1 user-defined and 3 built-in scattering model. The listing below then calculates the transport properties for each user defined temperature and calculated chemical potential using the RTA for user defined scattering models. It outputs a file containing the electronic conductivities ($\sigma$) that were used to compare to experiments. The code for the fitting procedure described in Appendix 2 is shown in the Listing 3 and 4. This reproduces the fitted conductivity curve shown in the inset of Figure \ref{GaAs optimization 1} and is used to calculate the so called fitted scattering rates.

\begin{sexylisting}{{\tt main.py} - Transport coefficients for GaAs - Base models}

import numpy as np
from  PAOFLOW import  PAOFLOW
from  PAOFLOW.defs.TauModel import TauModel

def main():

   PAOFLOW =  PAOFLOW. PAOFLOW(savedir='GaAs.save', smearing=None, npool=1, verbose=True)
  arrays,attr =  PAOFLOW.data_controller.data_dicts()
   PAOFLOW.read_atomic_proj_QE()
   PAOFLOW.projectability()
   PAOFLOW.pao_hamiltonian()
   PAOFLOW.interpolated_hamiltonian(nfft1=100, nfft2=100, nfft3=100)
   PAOFLOW.pao_eigh()
   PAOFLOW.gradient_and_momenta()

  doping = -3.5e17
   PAOFLOW.doping(tmin=380, tmax=812, nt=28, emin=-36, emax=2, ne=5000, doping_conc=doping)

  me = 9.10938e-31 # Electron Mass
  ev2j = 1.60217662e-19 # Electron Charge
  
  def acoustic_model ( temp, eigs, params ):

    from scipy.constants import hbar
    temp *= ev2j
    E = eigs * ev2j # Eigenvalues in J
    v = 5.2e3 # Velocity in m/s
    rho = 5.31e3 # Mass density kg/m^3
    ms = .7 * me #effective mass tensor in kg
    D_ac = 7 * ev2j # Acoustic deformation potential in J
    return (2*ms)**1.5*(D_ac**2)*np.sqrt(E)*temp/(2*np.pi*rho*(hbar**2*v)**2)

  acoustic_tau = TauModel(function=acoustic_model)

  fname = 'doping_n%s.dat'%np.abs(doping)
  temp = np.loadtxt('output/%s'%fname, usecols=(0,))
  mu = np.loadtxt('output/%s'%fname, usecols=(1,))

\end{sexylisting}
\begin{sexylisting}{{\tt main.py - continuation} - Transport coefficients for GaAs - Base models}

  channels = [acoustic_tau, 'polar_optical', 'impurity', 'polar_acoustic']

  tau_params = {'doping_conc':-3.5e17, 'D_ac':7., 'rho':5.31e3,
                  'a':5.653e-10, 'nI':3.5e17, 'eps_inf':11.6, 'eps_0':13.5,
                  'v':5.2e3, 'Zi':1, 'hwlo':[0.03536], 'D_op':3e10, 'Zf':6,
                  'piezo':0.16, 'ms':0.7, 'Ef':0.0}

  rho = []
  for t,m in zip(temp,mu):
    if  PAOFLOW.rank == 0:
      print('\nTemp, Mu: %f, %f'%(t,m))

     PAOFLOW.transport(tmin=t, tmax=t, nt=1, emin=m, emax=m, ne=1, scattering_channels=channels, tau_dict=tau_params, save_tensors=True, write_to_file=False)

    sigma = np.sum([sig for sig in np.diag(arrays['sigma'][:,:,0])])/3
    rho.append(1e2/sigma)

  if  PAOFLOW.rank == 0:
    with open('output/rho_rta_n3.5e17.dat' ,'w') as rho_file:
      for i,t in enumerate(temp):
        rho_file.write('%8.2f %9.5e\n'%(t,rho[i]))

   PAOFLOW.finish_execution()

if __name__== '__main__':
  main()
  
\end{sexylisting}

\begin{sexylisting}{{\tt main.py} - Transport coefficients for GaAs - Fitted models}
from  PAOFLOW import  PAOFLOW
from  PAOFLOW.defs.TauModel import TauModel
import numpy as np
import scipy.optimize
import sys

def main():

   PAOFLOW =  PAOFLOW. PAOFLOW(savedir='GaAs.save', smearing=None, npool=1, verbose=True)
  arrays,attr =  PAOFLOW.data_controller.data_dicts()
   PAOFLOW.read_atomic_proj_QE()
   PAOFLOW.projectability()
   PAOFLOW.pao_hamiltonian()
   PAOFLOW.interpolated_hamiltonian(nfft1=100, nfft2=100, nfft3=100)
   PAOFLOW.pao_eigh()
   PAOFLOW.gradient_and_momenta()

  def get_curve_eqn(x_data,y_data,x,degree):
     a  = np.polyfit(x_data,y_data,degree)
     a = np.array(a[::-1])
     curve_eqn = sum(a[j]*np.power(x,j) for j in range(len(a)))
     return curve_eqn

  def data_bin(unbinned_data,bin_size):
     binned_data = []
     for i in range(0,len(unbinned_data)-bin_size+1):
        binned_data.append(unbinned_data[i:i+bin_size])
     return np.array(binned_data)

  def cost_func(par,temp,mu,y_expt):
     y_calc = cost_func_driver(temp,mu,par)
     y_err = np.sum((y_calc-y_expt)**2)/len(y_expt)
     return y_err

  def cost_func_driver(temp,mu,par):
     pao_rho_list = []
     for t,m in zip(temp,mu):
     	 PAOFLOW.transport(tmin = t,tmax = t,nt = 1,emin=m, emax=m,ne = 1,scattering_channels=['polar_optical','impurity','polar_acoustic','acoustic'],tau_dict={'doping_conc':-3.5e17,'Ef':m,'D_ac':7.,'rho':5.31e3,'a':5.653e-10,'nI':3.5e17,'eps_inf':11.6,'eps_0':13.5,'v':5.2e3,'Zi':1,'hwlo':[0.03536],'D_op':3e10,'Zf':6,'piezo':0.16,'ms':0.291},a_imp=par[0],a_ac=par[1],a_pop=par[2],a_pac=par[3],write_to_file=False) 
     	pao_sigma = (arrays['sigma'][0,0]+arrays['sigma'][1,1]+arrays['sigma'][2,2])/3
        pao_rho = (1e2/pao_sigma) #convert to match units of expt data, ohm-cm
        pao_rho_list.append(pao_rho)
     return np.array(pao_rho_list)


  def cost_func_optimize(par_guess,bounds,temp,mu,y_expt):
     par_optimal = scipy.optimize.minimize(cost_func,par_guess,args=(temp,mu,y_expt),method='SLSQP',bounds=bounds,tol=0.0009)
     return par_optimal.x
\end{sexylisting}
\begin{sexylisting}{{\tt main.py - continuation} - Transport coefficients for GaAs - Fitted models}
  x_expt = np.loadtxt('expt_data/amith_n3.5e17',usecols=(0,)) 
  y_expt = np.loadtxt('expt_data/amith_n3.5e17',usecols=(1,))
  temp = np.loadtxt('EvT_n3.5e17/dope_TvsE_-3.5e+17.dat',usecols=(0,))
  mu = np.loadtxt('EvT_n3.5e17/dope_TvsE_-3.5e+17.dat',usecols=(1,))
  y_expt_curve=[]

  for t in temp:
      y_expt_curve.append(get_curve_eqn(x_expt,y_expt,t,4))  
      
  binned_temp = data_bin(temp,5)
  binned_mu = data_bin(mu,5)
  binned_y_expt = data_bin(y_expt_curve,5)

  for bin_no in range(len(binned_temp)):
      optimized_rho = open('optimized_rho_n3.5e17_bin_no_%s.dat'%bin_no,'w')
      optimized_S = open('optimized_S_n3.5e17_bin_no_%s.dat' %bin_no,'w')
      optimized_kappa = open('optimized_kappa_n3.5e17_bin_no_%s.dat' %bin_no,'w')
      if bin_no in range(1):
          par_guess = [1., 1., 1.,1.]
          bounds=((1e-9,100),(1e-9,100),(1e-9,100),(1e-9,100))
      else:
          par_guess = par_optimized
          bounds=((0.5*par_guess[0],1.5*par_guess[0]),(0.5*par_guess[1],1.5*par_guess[1]),(0.5*par_guess[2],1.5*par_guess[2]),(0.5*par_guess[3],1.5*par_guess[3]))
      par_optimized = cost_func_optimize(par_guess,bounds,binned_temp[bin_no,:],binned_mu[bin_no,:],binned_y_expt[bin_no,:])

      for t,m in zip(binned_temp[bin_no,:],binned_mu[bin_no,:]):
           PAOFLOW.transport(tmin = t,tmax = t,nt = 1,emin=m, emax=m,ne = 1,scattering_channels=['polar_optical','impurity','polar_acoustic','acoustic'],tau_dict={'doping_conc':-3.5e17,'Ef':m,'D_ac':7.,'rho':5.31e3,'a':5.653e-10,'nI':3.5e17,'eps_inf':11.6,'eps_0':13.5,'v':5.2e3,'Zi':1,'hwlo':[0.03536],'D_op':3e10,'Zf':6,'piezo':0.16,'ms':0.291},a_imp=par_optimized[0],a_ac=par_optimized[1],a_pop=par_optimized[2],a_pac=par_optimized[3],write_to_file=False)
          pao_sigma = (arrays['sigma'][0,0]+arrays['sigma'][1,1]+arrays['sigma'][2,2])/3
          pao_rho = (1e2/pao_sigma) #convert to match expt, ohm-cm
          pao_S = (arrays['S'][0,0]+arrays['S'][1,1]+arrays['S'][2,2])/3
          pao_kappa = (arrays['kappa'][0,0]+arrays['kappa'][1,1]+arrays['kappa'][2,2])/3
          optimized_S.write('%8.2f %9.5e\n'%(t,pao_S))
          optimized_kappa.write('%8.2f %9.5e\n'%(t,pao_kappa))
          optimized_rho.write('%8.2f %.4f %.3f %.3f %.3f %.3f %9.5e\n'%(t,m,par_optimized[0],par_optimized[1],par_optimized[2],par_optimized[3],pao_rho))

      optimized_rho.close()
      optimized_S.close()
      optimized_kappa.close()

   PAOFLOW.finish_execution()

if __name__== '__main__':
  main()

\end{sexylisting}


\newcommand{\Ozolins}{Ozoli\c{n}\v{s}}

\end{document}